\documentstyle[11pt]{article}

\begin{document}

		\begin{titlepage}
\setcounter{page}{1}

\date{}
\title{\small\centerline{July 1993 \hfill NUHEP-TH-93-17}
\small\rightline{UPR-0576T}
\small\rightline{DOE-ER\,40757-019}
\small\rightline{CPP-93-19}
\bigskip\bigskip
{\LARGE\bf Loop-induced majoronic decay of neutrinos}\bigskip}

\author{Darwin Chang\\
\normalsize \em Department of Physics, Northwestern University,
Evanston, IL 60208, USA
\medskip \\
Jiang Liu \\
\normalsize \em Department of Physics, University of Pennsylvania,
Philadelphia, PA 19104, USA
\medskip\\
Palash B. Pal\\
\normalsize \em Center for Particle Physics, University of Texas,
Austin, TX 78712, USA}

\maketitle
\vfill
		\begin{abstract}\noindent\normalsize
In models where the suppression of the tree level couplings of
fermions with Goldstone bosons is not protected by symmetry, the
loop-induced couplings can in general dominate over the tree
couplings. We demonstrate this by calculating the decay $\nu\to \nu'
+ \mbox{Majoron}$ in the simplest singlet Majoron model. It is shown
that 1-loop contributions can be orders of magnitude bigger than the
tree-level result, and hence allows a wide range of interesting values
of neutrino mass and mixing to be compatible with cosmological constraints.
		\end{abstract}
\vfill

\thispagestyle{empty}
		\end{titlepage}


Theoretical models with $B-L$ as a spontaneously broken global
symmetry have a Nambu-Goldstone boson called Majoron ($J$),
where $B$ and $L$ are respectively the baryon and lepton number.  In
these models a neutrino can decay to  a Majoron,
$\nu\to\nu' + J$.
Such an invisible decay can be of cosmological interest. In
particular, a neutrino with a mass lying in the cosmologically
excluded region between 30 eV and 2 GeV for stable neutrinos \cite{CM}
would still be allowed if it decays sufficiently fast.  In that case
the standard cosmological constraint derived from the mass density of the
universe is \cite{DKT}
	\begin{eqnarray}
m_{\nu}\sqrt{\tau_\nu/t_0} < 30 \, {\rm eV},  \label{densitybd}
	\end{eqnarray}
where $m_{\nu}$ and $\tau_\nu$ are the mass and the lifetime of $\nu$,
whereas $t_0 \approx 5 \times 10^{17}$ s is the present age of the
universe.  The bound can be strengthened to \cite{StTu}
	\begin{eqnarray}
m_\nu \sqrt{\tau_{\nu}/10^4\, {\rm s}} < 1\, {\rm MeV},
\label{galaxybd}
	\end{eqnarray}
provided one assumes the canonical picture of galaxy
formation.  However, since the physics of galaxy formation is not
fully understood, this bound should be taken less seriously than the
previous one.

In many simple versions of Majoron Models the decay process
occurs with a negligibly small rate \cite{fastdecay} at the tree
level.  This follows because to a first approximation the
neutrino-Majoron coupling matrix often turns out to be proportional to
the neutrino mass matrix \cite{ScVa}.  Thus, in the mass
eigenstate basis the neutrino-Majoron interaction is  diagonal and the
decay wouldn't occur at this level. To have a nonzero
tree-level off-diagonal coupling it is necessary to consider
non-leading terms, which are related to the mixings of the light
neutrinos $\nu$ and $\nu'$ with some other very heavy species and
hence highly suppressed.

This feature is generally unprotected by symmetries, however.
Modifications introduced by loop corrections
can therefore be potentially significant, leading
to a rare situation in which the one loop
correction is larger than the tree-level result.
Consider the earliest Majoron model of Chikashige, Mohapatra and Peccei
\cite{CMP} as an example.  The model is simple in the sense that it
requires only minor modifications over the standard model. Therefore
the calculation is relatively simple,  and the result is dramatic.
The model differs
from the standard model by a right-handed neutrino $N_R$ for
each generation, and  a Higgs singlet
	\begin{eqnarray}
\Phi(x) ={1\over \sqrt 2} (v_1 + s(x) +iJ(x)),  \label{singlet}
	\end{eqnarray}
which carries 2 units of $B-L$ quantum number.  The vacuum
expectation value of $\Phi$, $v_1$, breaks $B-L$
spontaneously and gives rise to a Majoron $J$.  As
a gauge singlet, $J$ does not interact directly with the $Z$ boson,
and hence unlike many other later versions this model has
not been ruled out by the measurements of $Z$-width.

At the tree level, the majoronic decay of neutrino  is determined by
the Yukawa interaction
	\begin{eqnarray}
 -{\cal L}_Y =\sum_{a,A}
{\sqrt 2\over v_2} {\cal D}_{aA} \overline \psi_{La} \phi N_{RA}
+ \sum_{A,B} {1\over \sqrt 2v_1} {\cal M}_{AB} \overline {\widehat{N}}_{LA}
\Phi N_{RB}  + \mbox{h.c.,}
  \label{yukawa}
	\end{eqnarray}
where $\psi_L$ is the usual lepton doublet and
$\widehat N = C\gamma_0 N^*$.  $\phi$ is the standard Higgs doublet
with its neutral component $\phi^0$ given by
	\begin{eqnarray}
\phi^0 (x) = {1\over\sqrt 2}(v_2 + H(x) + i\varphi(x)).
\label{doublet}
	\end{eqnarray}
$\cal D$ and $\cal M$ are coupling matrices in flavor space, $v_2$ is
the vacuum expectation value which breaks the gauge symmetry.  The
mass terms of the neutrinos are given by
	\begin{eqnarray}
- {\cal L}_{\rm mass} = {1\over2} \left( \begin{array}{cc} \overline
\nu_L \quad \overline{\widehat N}_L  \end{array} \right)
\left( \begin{array}{cc} 0  &  {\cal D}  \\
                 {\cal D}^T &  {\cal M}   \end{array} \right)
 \left( \begin{array}{c} \widehat\nu_R \\ N_R
\end{array} \right) + \mbox{h.c.},
  \label{massmatrix}
	\end{eqnarray}
where each entry in the mass matrix represents a $3\times 3$ matrix
for 3 generations of fermions.

In the present model the elements  of $\cal D$ arise  from
the same physics that gives rise to the masses of charged
leptons.  As a result, we assume
	\begin{eqnarray}
{\cal D}_{aA} \ll gv_2 \qquad \forall a,A.
\label{D<<v2}
        \end{eqnarray}
where $g$ is the SU(2) gauge coupling.
To proceed, we  further assume that CP is conserved and that
	\begin{eqnarray}
r \equiv v_2/v_1 \ll 1 \,.
\label{r<<1}
	\end{eqnarray}
This implements the well-known see-saw mechanism \cite{seesaw} for the
generation of small neutrino masses.  In addition, we assume for
simplicity that
	\begin{eqnarray}
{\cal M}_{AB} \sim v_1  \qquad \forall A,B.
	\end{eqnarray}
These assumptions are not crucial. However they  do simplify the
calculation.

Calling the doublet Yukawa couplings collectively as $y$, we can then
diagonalize the mass matrix perturbatively in powers of $r$ and $y$.
To order $r^2y^2$ one obtains
the following relations between the mass eigenstates and
the weak eigenstates \cite{Kan80}:
	\begin{eqnarray}
\left( \begin{array}{c} \nu_L\\  \widehat{N}_L \end{array} \right) &=&
\left( \begin{array}{cc} 1 - {1\over2} \rho \rho^T & \rho \\
- \rho^T  & 1 - {1\over2} \rho^T \rho
\end{array} \right)
\left( \begin{array}{cc} f^* & 0 \\  0 & F^*
\end{array} \right)
\left( \begin{array}{c} \chi_L \\  X_L \end{array} \right), \nonumber
\\
\label{diagonalization}
\\
\left( \begin{array}{c} \widehat\nu_R\\  N_R \end{array} \right) &=&
\left( \begin{array}{cc} 1 - {1\over2} \rho \rho^T & \rho \\
- \rho^T  & 1 - {1\over2} \rho^T \rho
\end{array} \right)
\left( \begin{array}{cc} f & 0 \\  0 & F \end{array} \right)
\left( \begin{array}{c} \chi_R \\  X_R \end{array} \right), \nonumber
	\end{eqnarray}
where $\rho \equiv {\cal DM}^{-1} \sim ry$.  To order $\rho^2$, the
mass terms are
\begin{eqnarray}
- {\cal L}_{\rm mass} = {1\over2} \left[ \overline \chi_L m \chi_R +
\overline X_L M X_R \right] + \mbox{h.c.},
	\end{eqnarray}
with diagonal mass matrices
	\begin{eqnarray}
m &=& f^T \left( - \rho {\cal M} \rho^T \right) f, \label{m}\\
M &=& F^T \left( {\cal M} + {1\over2} \rho^T \rho {\cal M} + {1\over2}
{\cal M} \rho^T \rho \right) F.
  \label{M}
	\end{eqnarray}
The diagonalization into a light block and a heavy
block is performed by the first matrix on the right hand side of Eq.
(\ref{diagonalization}). The two unitary matrices $f$ and $F$
determine the mixings among the light-light and the heavy-heavy
neutrinos, respectively.  Notice
$f$ and $F$ can be complex, even if CP is conserved, unless
the Majorana eigenstates all have the same CP eigenvalue \cite{Wol81}.

{}From Eqs. (\ref{yukawa}), (\ref{diagonalization})
and (\ref{m}) one
can  easily show that, to ${\cal O}(\rho^2) \sim {\cal O}(r^2y^2)$,
the $\chi\chi J$ coupling matrix is proportional to $m$ which is
diagonal \cite{ScVa}.  Although the next order terms are not diagonal
\cite{MoPa88}, they are ${\cal O}(\rho^4) \sim {\cal O}(r^4y^4)$,
rendering their contributions
to the neutrino decay too small to be of practical interest.
We shall show that this feature is modified drastically by
radiative corrections.   To discuss this effect, let us
parameterize the effective interaction for the decay
$\chi_a\to \chi_b + J$, keeping terms upto ${\cal
O}(r^2y^2)$, as
	\begin{eqnarray}
{\cal{L}}_{{\rm eff}}=
- i \overline \chi_b \left(
{m_a \over v_1} \gamma_5 \delta_{ba} +
{\cal T}_{ba}R - {\cal T}^*_{ba} L \right) \chi_a J +
\mbox{h.c.},  \label{Tab}
	\end{eqnarray}
in which the first term is the tree-level result and
${\cal T}_{ba}$ comes from loop contributions to be
calculated below. The 1-loop diagrams for the decay  in the Feynman
gauge are shown in Figs.~\ref{f:ZHphi}, \ref{f:J+s}, and \ref{f:Js}.
Graphs with an $H$-$s$ mixing are negligible due to  (\ref{r<<1}).
Charged-current interactions do not make any contribution at this
level.  The Feynman rules for relevant couplings are
summarized in Fig.~\ref{f:Feynman}.

The contributions of various diagrams are
organized in  accordance with the virtual bosons propagating in the
loop.  For example, ${\cal T}^{(Z)}$ represents the contribution
from all diagrams containing a virtual  $Z$.
Neglecting the mass of the
final state neutrino and using the shorthanded
	\begin{eqnarray}
P = f^T {\cal DM}^{-1} F^* \,,
\label{P}
	\end{eqnarray}
we find,
upto terms of order ${\cal O}(r^2y^2)$ times a potential logarithmic factor,
	\begin{eqnarray}
{\cal T}_{ba}^{(Z,H,\varphi)} &=& 0, \label{ZHphi}\\
 {\cal T}_{ba}^{(s)}  &=& {-1\over 8\pi^2}
\sum_{A} P_{bA}P_{aA}  \left({M_{A}\over v_1} \right)^3
           I(M^{2}_{A}, M_s^2), \label{s} \\
 {\cal T}_{ba}^{(J)}  &=&
{1\over 8\pi^2}\sum_{A} P_{bA}P_{aA}
 \left({M_{A}\over v_1} \right)^3 I(M^{2}_{A}, 0), \label{J}\\
 {\cal T}_{ba}^{(Js)}  &=&
{1\over 8\pi^2}\sum_{A} P_{bA}P_{aA}
 \left({M_{A}\over v_1} \right)^3
{M_s^2\over M_{A}^2-M_s^2}\ln{M_{A}^2\over M_s^2} \,, \label{Js}
	\end{eqnarray}
where $M_s$ is the mass of the singlet scalar $s$  and
	\begin{eqnarray}
I(x,y)= \int_0^\infty d\zeta {\zeta \over (\zeta+x)(\zeta+y)}
	\end{eqnarray}
which is divergent. However, the sum
of  Eqs. (\ref{ZHphi}-\ref{Js}) is finite:
	\begin{eqnarray}
{\cal T}_{ba} = 2 {\cal T}_{ba}^{(Js)} .  \label{14}
	\end{eqnarray}

Using the definition of $P$ from Eq. (\ref{P}),  one sees that ${\cal
T}_{ba}$ is of order $m_a/4\pi^2v_1$ times an interfamily mixing for
$M_A\sim v_1$ and  reasonable choices of $M_s/M_A$.  Despite the
loop suppression factor $1/4\pi^2$, for a wide range of interesting
values of $m$ and $M$ this result is orders of magnitude bigger than
the tree-level result, which is ${\cal O}(r^4y^4)$, as pointed out
earlier.

There is a technical subtlety associated
with the choice of the representation of Majoron fields.  Instead of
using Eq. (\ref{singlet}) as the definition of $J$, one can use a
non-linear representation:
	\begin{eqnarray}
\Phi(x)={1\over \sqrt 2}(v_1+s(x)) e^{iJ(x)/v_1}.
\label{angular}
	\end{eqnarray}
The interaction Hamiltonian now  has momentum-dependent terms.  As a
result, the formulation
of Feynman rules is different \cite{GJLW} from the usual situation.
The two representations, Eqs. (\ref{singlet}) and (\ref{angular}), give
identical results for tree-level processes in which momentum-dependent
interactions are negligible.  For the case at hand, a naive
calculation following (\ref{angular}) without taking into account the
associated Feynman rule difference would incorrectly give ${\cal
T}_{ba}^{(Js)}=0$, and as a result the final answer would be a factor of
2 smaller.

Let us now examine the significance of the loop result.
With our assumption of CP invariance, the mixing matrices satisfy the
constraints \cite{Wol81,Kay84}
$f_{ab}^* = \eta_b f_{ab}$,
and a similar one for $F_{AB}$, where the CP eigenvalue of $\chi_b$
is $i\eta_b$, with $\eta_b=\pm 1$. Thus,
	\begin{eqnarray}
P_{aA}^* = \eta_a \eta_{A} P_{aA} \,.
	\end{eqnarray}
Due to the Majorana nature of the neutrinos, the hermitian conjugate
term in Eq. (\ref{Tab}) contributes equally as the other one.
The amplitude of the decay can therefore be written as
	\begin{eqnarray}
{\cal{A}}(\chi_a\to \chi_b + J)= {1 \over 2\pi^2}
\theta_{ba} \overline u_b (R - \eta_a \eta_b L) u_a,
\label{amplitude}
	\end{eqnarray}
where $u_{a,b}$ are the standard spinors,
and for convenience we have defined a parameter $\theta_{ba}$:
	\begin{eqnarray}
\theta_{ba} \equiv \sum_{A}
P_{bA}P_{aA} \left({M_{A}\over v_1} \right)^3
{M_s^2\over M_{A}^2-M_s^2}\ln{M_{A}^2\over M_s^2},  \label{theta}
	\end{eqnarray}
which plays the role of $m_a/v_1$ times an effective
light-light interfamily mixing.  Note that $\theta$ vanishes when
$v_1\to\infty$, since the heavy neutrinos decouple in that limit.
Eq. (\ref{amplitude}) demonstrates that the transition amplitude is
purely pseudoscalar if the two Majorana states have the same CP
property, and purely scalar if their CP eigenvalues are opposite.
Intuitively, this is obvious since the CP eigenvalue of the Majoron is
negative.

{}From Eq. (\ref{amplitude}), the rate of the decay for either sign of
$\eta_a\eta_b$ is
	\begin{eqnarray}
\tau^{-1}(\chi_a\to \chi_b J) = {\theta_{ba}^2 m_a \over 32\pi^5} \,.
\label{rate}
	\end{eqnarray}
Substituting (\ref{rate}) into (\ref{densitybd}) yields
	\begin{eqnarray}
\theta_{ba}^2 (1\, {\rm eV}/m_a) > 1.5 \times 10^{-32} \,.
\label{bound}
	\end{eqnarray}
This result implies that, for a wide range of $v_1$ and
neutrino masses and mixings, the radiatively-induced neutrino-Majoron
decay is fast enough to allow the neutrino be compatible with the big
bang cosmology.

To have a feeling for
our result, we take from  (\ref{diagonalization}), (\ref{m})
and (\ref{theta}) our order-of-magnitude estimates $m_a \sim v_1
\rho^2$ and $\theta \sim \rho^2$.  Replacing $\rho$ by $m_D/v_1$,
where $m_D$ is the relevant Dirac mass, we obtain
$m_a \sim  10^4\,{\rm eV} \times (1\,{\rm TeV}/v_1) \times (m_D /
100\, {\rm MeV})^2$.  Thus for $v_1 = 1\, {\rm TeV}$ and  $m_D =
m_\mu$, $m_a\approx10\, {\rm keV}$. Such a neutrino can easily satisfy
the cosmological constraint (\ref{densitybd}) which only  requires from
(\ref{bound})
	\begin{eqnarray}
m_a  > 1.5 \cdot 10^{-8}\, {\rm eV} \,\times (v_1/1\, {\rm TeV})^2.
\label{bound2}
	\end{eqnarray}
For $v_1$ of order TeV, even after taking all the uncertainties of an
order-of-magnitude estimate, this shows that the mechanism can provide
a fast decay for neutrinos of any mass.
One can also express Eq. (\ref{bound2}) as a bound on $v_1$, i.e.,
$v_1 < 10^4 \,{\rm TeV} \times (m_D/ 100\,{\rm MeV})^{2/3}$.
If one uses the bound  of galaxy formation from Eq. (\ref{galaxybd}),
the constraint is
similar to Eq. (\ref{bound}) with the right hand side replaced by
$6.6\times 10^{-28}$.  The corresponding bound on the neutrino mass is
$m_a > 6.6 \cdot 10^{-4} \,{\rm eV}\, \times (v_1/1\,{\rm TeV})^2$
with the same phenomenological conclusion.

On the other hand, if one has to rely on the tree level contribution
only \cite{MoPa88}, instead of $\theta/2\pi^2$ we obtain $2\rho^4$ as
the order-of-magnitude estimate for the  coefficient in the amplitude
of Eq. (\ref{amplitude}), the factor 2 occuring due to the Majorana
nature of the neutrinos.   The bound in Eq. (\ref{densitybd}) can then
be satisfied if
	\begin{eqnarray}
m_a > 20\,{\rm keV} \times \left( {v_1/ 1\,{\rm TeV}} \right)^{4/3} \,.
	\end{eqnarray}
This can be satisfied for a narrow range of values of the mass of
 $\nu_\mu$ provided $v_1$ is not much larger than a TeV. However, for
the bound in Eq.
(\ref{galaxybd}), the factor 20 keV from this equation should be
replaced by 750 keV, which cannot be satisfied for any $\nu_\mu$ mass
allowed by experiments.

In conclusion, we have calculated 1-loop radiative corrections to
the neutrino-Majoron decay in a singlet Majoron model. The result
shows that loop contributions can decisively dominate the tree-level
contribution,  allowing  a wide range of interesting values
of neutrino mass and mixing be compatible with the big bang cosmology.
Note that the sole reason that the 1-loop result dominates over the tree
level result is just because the tree level suppression is accidental
in nature, that is, it is not enforced by any symmetry.  It has little to do
with the other details of the model.  Therefore, one expects that the same
conclusion can be drawn in a large class of models in which there is
such accidental tree level suppression of  off-diagonal majoron
couplings.  In addition,  since the accidental suppression disappears
at the 1-loop, one does not expect that including higher loop
contributions would change our result qualitatively.

\paragraph*{Acknowledgements~:}
This work was supported in part by the U.S. Department of Energy (D.C.
and P.B.P.), and an SSC Fellowship (J.L.) from Texas National Research
Laboratory Commission. We thank L. Wolfenstein, L.~F. Li, N.~G.
Deshpande, D.~A. Dicus and  S. Paban for discussions at various stages
of this work.

	
\bigskip \bigskip \bigskip
\centerline{\bf Figure Captions}
		\begin{figure}[h]
\vspace{5mm}
\caption[]{One-loop graphs for the leading terms of the decay
$\chi_a\to \chi_b+J$ involving a $Z$, $\varphi$ or $H$ internal line.
The contribution of these diagrams appear in Eq. (\ref{ZHphi}).}
\label{f:ZHphi}
%
\vspace{5mm}
\caption[]{Same as Fig.~\ref{f:ZHphi} except the internal boson lines
are either $s$ or $J$. The contribution of these diagrams are
summarized in Eqs. (\ref{s}) and (\ref{J}).}
\label{f:J+s}
\vspace{5mm}
\caption[]{Diagrams involving the trilinear scalar vertex $JJs$ which
contribute at the leading order. Their contribution is
summarized in Eq. (\ref{Js}).}
\label{f:Js}
\vspace{5mm}
\caption[]{Feynman rules relevant for the calculation of
1-loop corrections to the decay $\chi_a\to\chi_b+J$.
The Feynman rules for fermion vertices with $H\,(s)$ can be obtained
by multiplying the corresponding $\varphi\,(J)$ vertex by
$-i\gamma_5$.}
\label{f:Feynman}
		\end{figure}

\end{document}